\documentclass[traditabstract]{aa} 
\usepackage[utf8]{inputenc}

\usepackage{graphicx}
\usepackage{natbib}
\usepackage{longtable,lscape}
\usepackage{rotating}

\usepackage{txfonts}
\newcommand{\msun}{\mbox{$\rm M_\odot\,$}}

\newcommand{\lsun}{\mbox{$\rm L_\odot\,$}}

\newcommand{\mum}{\mbox{$\rm \mu m\,$}}

\def\arcmin{\hbox{$^\prime$}}
\def\arcsec{\hbox{$^{\prime\prime}$}}
\def\kms{km s$^{-1}$}

\begin{document}

\title{Radiatively driven Rayleigh-Taylor instability candidates
  around a forming massive star system}

\subtitle{NACO adaptive optics and VISIR study of G333.6-0.2}
\titlerunning{Radiative Rayleigh Taylor instability candidates in
  G333.6-0.2 }

   \author{M. S. N. Kumar\inst{1}
            }

   \institute{Centro de Astrofísica da Universidade do Porto,
              Rua das estrelas, 4150-762 Porto, Portugal\\
              \email{nanda@astro.up.pt}
  }


 \abstract{}
 
 \abstract {The formation of the highest mass stars are thought to be
   dominated by instabilities resulting from gravitation and
   radiation. Instabilities due to gravitation are commonly
   demonstrated by observations of fragmentation, but those due to
   effects of radiation have thus far not been found.  Here I report
   on the NACO adaptive optics and mid-infrared diffraction-limited
   VISIR imaging data of an extemely luminous ultra-compact HII region
   G333.6-0.2. Two infrared sources, one bright in the near-infrared
   (appearing point-like) and another in the mid-infrared (resolved
   with an elliptical shape) are uncovered through this data, which
   are located at the heart of this region. These infrared sources
   appear to be embedded in the waist of a bipolar-shaped nebula and
   UCHII region, the lobes of which are separated by a dark
   patch. Dense filamentary features with finger/hook morphology are
   found; they appear to be connected to the two bright infrared
   sources and the sizes of these hook features are {\em sharply
     limited} to $<$5000~AU. The observed properties of this target
   and a large amount of previous data obtained from the literature
   are compared together with the results of various numerical
   simulations of high-mass star formation. This comparison favours
   the interpretation that the finger/hook-like structures likely
   represent radiatively driven Rayleigh-Taylor instabilities arising
   in the outflow cavity of a forming high-mass binary star system.}



   \maketitle
%

\section{Introduction}
The formation of the most massive stars in our Galaxy is an enigmatic
topic in astrophysics revolving around a singular problem: the
dissipation of intense radiation pressure in the early stages of
massive protostellar collapse, which is thought to be capable of
reversing accretion flows \citep[see][]{zy07}. The work-around
solution to this problem suggested by theory argues that stars as
massive as 140~\msun~ can form through disk-mediated accretion
\citep{nakano89,ys02,mk09,rk10} because radiation is dissipated
through outflow cavities.  In recent years, detailed numerical
simulations of massive star formation have been carried out; the
majority of these are radiation-hydrodynamic
\citep{mk09,rk10,ajc11,rk11,peters10a,peters10b}, with an increasing
focus on magneto-hydrodynamic simulations \citep{hen11,myers13}. Each
of these simulations varies in its initial conditions and focus,
leading to different approximations and limitations impacting the way
in which realistic systems are represented. Most importantly,
simulations using adaptive mesh refinement codes together with
non-axisymmetric geometries \citep[e.g.][]{mk09,hen11,peters10a}
produce distinct fine structures at physical scales of a few hundred
to a thousand AU.

For example, radiation hydrodynamic simulations of \citet[][hereafter
  KKM09]{mk09}, and \citet{rk10,rk11} show distinctly different
features due to such treatment. While the radiation-pressure-driven
bubble rises along the outflow axis in both cases, KKM09 simulations
produced Rayleigh-Taylor instabilities (RTI) in the outflow bubble
where radiation itself behaves like a lighter fluid penetrating into
the denser gaseous matter (heavy fluid) surrounding the star. The
optically thin gaps in between the RTI allowed radiation to escape,
and the overdense material falls back to the disk plane. On larger
scales, infalling gas is thought to strike the walls of the outflow
cavity bubble, which effectively slides and settles down to the disk
plane. The simulations of KKM09 also amplify the gravitational
instabilities built up due to the swing mechanism \citep{ars89},
resulting in two massive stars. However, they used grey flux-limited
diffusion (FLD) radiative transfer without including the direct
radiation pressure.  On the other hand, although \cite{rk10,rk11}
used frequency-dependent ray-tracing for direct radiation they did not
produce RTI features and thus eventually questioned their stability as
discussed in \cite{rk12}. The topic of observational investigation in
this work is therefore centred on the role of RTI in the mechanism of
high-mass star formation.

Observations of ultra-compact HII regions at sub-arcsecond resolutions
\citep{wc89,kcw94,urq09}, mostly from the Very Large Array, have thus
far not uncovered significant sub-structures at a level of a few
hundred to a thousand AU. For instance, adaptive optics observations
with the 8.2\,m Very Large Telescope (VLT) of the UCHII region
G5.89-0.39 \citep{feldt03} in the K$_s$ and L$^\prime$ bands reveal
structure at such small scales. However, those observations were
obtained much before recent theoretical developments allowed direct
comparison. Finding physical structures resulting from instabilities
associated with high-mass star formation depends not only on
high-quality observations but also on favourable targets that satisfy
the criteria of being nearby, young, and extraordinarily
luminous. Massive (OB- type) stars are very few, representing just
over 0.1\% of all main-sequence stars in our Galaxy.  Thus, stringent
conditions on observations and target selection exist in order to
verify the validity of theoretical scenarios.

The UCHII region G333.6-0.2 is known to possess some extraordinary
qualities and is situated inside the giant HII region complex RCW106,
which represents parts of a long filamentary giant molecular cloud
\citep{bm04}.  Soon after the advent of far-infrared astronomy in the
late 1960s, it was realized that this is a powerful source
\citep{becklin73}, comparable with the Galactic centre and the famous
stellar beast, Eta Carinae \citep{sutton74}. Located at a distance of
3.1 -- 3.6~kpc to Sun, it has a bolometric luminosity of
3$\times$10$^6$~\lsun\, and is diagnosed as being compact (11\arcsec\,
diameter) and dusty, as well as hosting highly ionizing stars
\citep{becklin73,natta85}.  A distance of 3.6~kpc will be adopted in
this work to allow conservative interpretation of the specific results
here. The observed Lyman continuum flux (LyC) from this compact region
is estimated to be N(LC)$\sim$ 1$\times$10$^{50}$~sec$^{-1}$, which is
equivalent to that emitted by about 19 O7V stars
\citep{fujiyoshi06}. The UCHII region itself is embedded in a massive
molecular clump of $\sim$15~000~\msun\, \citep{bm04} from which a
forest of emission lines tracing dense gas is detected
\citep{lo09}. The dense clump hosts an embedded cluster, but the
infrared and radio peaks within the target are coincident and
compact. At sub-arsec resolution, this compact peak reveals itself as
a dark patch in the near-infrared K band, which forms the waist of a
bipolar-shaped UCHII region and infrared nebula. The dark patch is
suggestive of a large toroidal structure, and the bipolar shaped HII
region and infrared nebula likely represent a radiation-driven outflow
bubble (see Figure.~1). Here I present adaptive optics observations of
this compact region and analyse it using various infrared and radio
data at (sub)arcsecond resolutions. This exercise has resulted in the
discovery of small-scale density structures ($<$1\arcsec) with
filamentary and finger/hook-like morphology associated with two
luminous sources that appear point-like at 0.1\arcsec-0.3\arcsec
level.  These data are used to confront some outcomes produced by
various numerical simulations.

\section{Observations and Data Reduction}

Data used in this work are extracted and processed from various archives.

{\em Infrared observations:} Near-infrared adaptive optics imaging
data in the J, H, K$_s$, Br$\gamma$ (NB2.17), and L$^{\prime}$ bands
were obtained using the NAOS-CONICA (NACO) system, which is mounted on
the European Southern Observatory (ESO) 8.2~m VLT Yepun. These data
correspond to the proposal ID: 075.C-0424(A)(PI: Sch$\ddot{o}$del).
Mid-infrared 12~$\mum$ and 18~$\mu$m imaging observations were made
using the VISIR/VLT (ESO proposal ID: 60.A-9234(A) PI: Siebernmorgen).
A broad-band K$_s$ band image from the VISTA Variables in the Via-Lactea
(VVV) survey \citep{minniti10} was retrieved to calibrate astrometry
and photometry of the adaptive optics data. The spatial resolution
achieved with NACO is highest for the L$^{\prime}$ band at 0.1\arcsec,
slightly lower in the J, H, K$_s$, and Br$\gamma$ bands between
0.1\arcsec-0.2\arcsec, and diffraction limited at 0.3\arcsec and
0.5\arcsec\, in the 12~$\mum$ and 18~\mum bands respectively.

\begin{figure}
\centering
    \resizebox{\hsize}{!}{\includegraphics{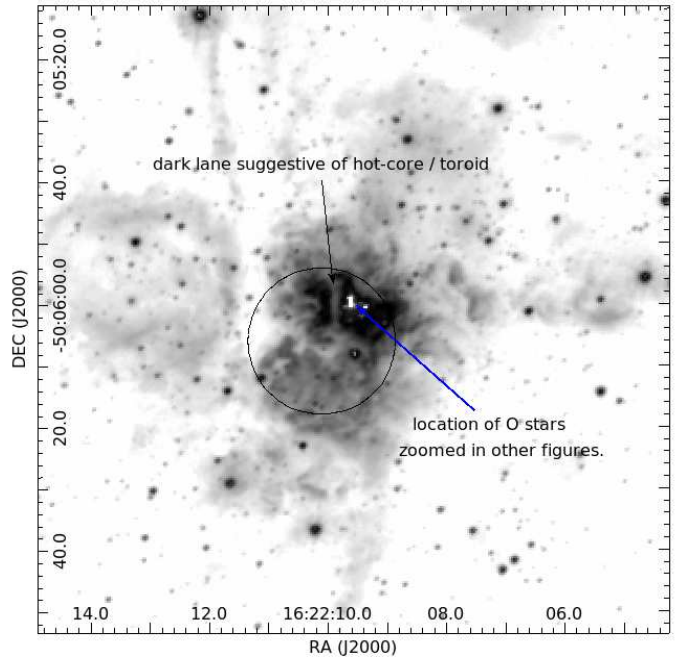}}
       \caption{A near-infrared Ks-band image of the source from the
         VVV survey (0.5\arcsec\, resolution) is shown using inverted
         grey scale and logarithmic intensity stretch. The circle
         marks the 24\arcsec beam of the 1.2~mm observations that
         uncovered the cold and dense (n$_{H_2}$=
         5.44$\times$10$^4$~cm$^{-3}$) molecular clump \citep{bm04},
         with an estimated mass of $\sim$15~000~\msun\, and a radius
         of 1.88~pc. Intense line emission from a variety of molecular
         species (CO, CS, HCO$^+$, HCN, HNC, C$_2$H) is detected from
         this clump \citep{lo09}. The bipolar-shaped nebula and the
         dark lane are identical in morphology to the UCHII region
         \citep{fujiyoshi06}. The dark lane (size $\sim$
           30~000~AU) may represent a hot core/toroid seen edge-on,
           but further analysis here indicates that it may actually
           represent the low-density part of the red-shifted outflow
           cavity (see Sec.5).}

\end{figure}

Individual frames were recalibrated and combined to produce mosaics in
each band. Astrometric calibration for images from J to L$^{\prime}$
band was achieved by a triangulation method using the position of all
stars in the NACO images against 2MASS positions. On the VISIR images,
the target appears mostly extended, without any well-defined point
source. Therefore we applied a cross-correlation of the nebular
features between VISIR images and calibrated the NACO
L$^{\prime}$-band image. The final astrometric accuracy is, therefore,
limited by the 2MASS accuracy of $\sim$0.1\arcsec\,. Aperture
photometry was performed for selected sources using apertures of
0.5\arcsec\,. Photometric zero-points measured at the VLT were applied
first. Subsequent improvement was achieved by cross-calibrating the
first pass photometry with that of VISTA photometry. Since the VISTA
photometry is tied to the 2MASS photometry using several hundred stars
in a given frame, it is robust with conservative maximum errors of
10\%.

{\em Chandra X-Ray data analysis:} Chandra X-Ray telescope
observations of this target were obtained through observation ID:9911
(PI: L. Townsley).  The data were reprocessed using CIAO software,
following the analysis threads on the Chandra website to produce event
files. The final goal is to evaluate the pressure from shocked gas
resulting from stellar winds (sec.~3.1). A background-subtracted
spectrum was extracted for a region covering the features of interest
without including point sources (stars). The background region was chosen
about 4\arcmin\, away from the G333.6-0.2 UCHII region. The spectrum
was fitted using two XSPEC components (phabs and apec). A
statistically good fit was obtained with a reduced chi-squared value
of 0.99, with 10 degrees of freedom. The resulting fit and the
  residuals are shown in Fig.~2 (electronic version only) For the
phabs model, we began with an initial guess of the line-of-sight
hydrogen column density of 2.17$\times$10$^{22}$~cm$^{-2}$. However,
the modelling results in a final value of n$_H$ =
4.46$\times$10$^{22}$~cm$^{-2}$; this is a consistent value,
considering that the UCHII region is embedded inside a dense and
massive molecular clump. The APEC model component resulted in an X-ray
temperature of {\em kT$_X$} = 2.48 keV. The emission measure was
obtained using the normalization parameter from the APEC model: norm =
$\frac{EM \times 10^{-14}}{4\pi D^2}$, where EM is the emission
measure, D is the distance to the source. Then the number density
n$_X$ was calculated using n$_x$=$\sqrt{\frac{EM}{V}}$, where V is the
volume of the region as measured from the region file used to extract
the X-ray spectrum. The size of the UCHII region (10\arcsec) is
considered to be the diameter of a sphere in computing the volume, and
a distance of 3.6~kpc to the target is assumed.

\begin{figure}
   \centering

 \resizebox{\hsize}{!}{\includegraphics{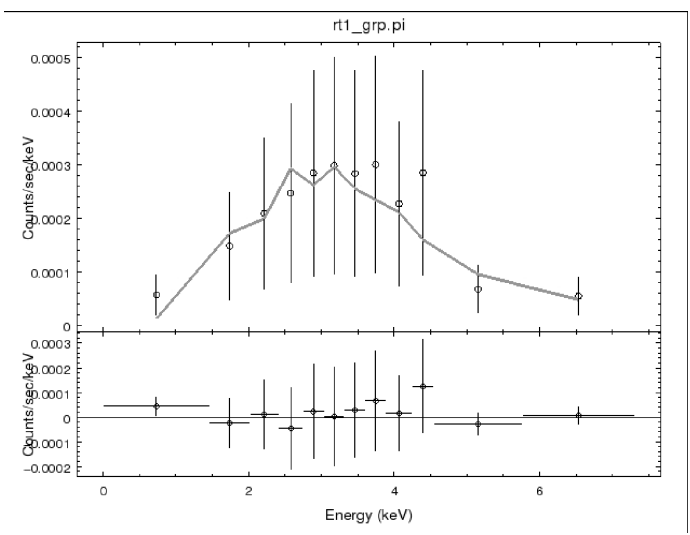}}
      \caption{ Chandra X-ray spectrum analysis. The top panel shows
        the X-ray spectrum for the region, including the IRS\,1 and
        IRS\,2 sources. The solid grey line shows the fitted XSPEC
        model with phabs and apec components. The bottom panel displays
        the residuals after fitting.  }
         \label{images}
   \end{figure}

\begin{figure*}
   \centering

 \resizebox{\hsize}{!}{\includegraphics{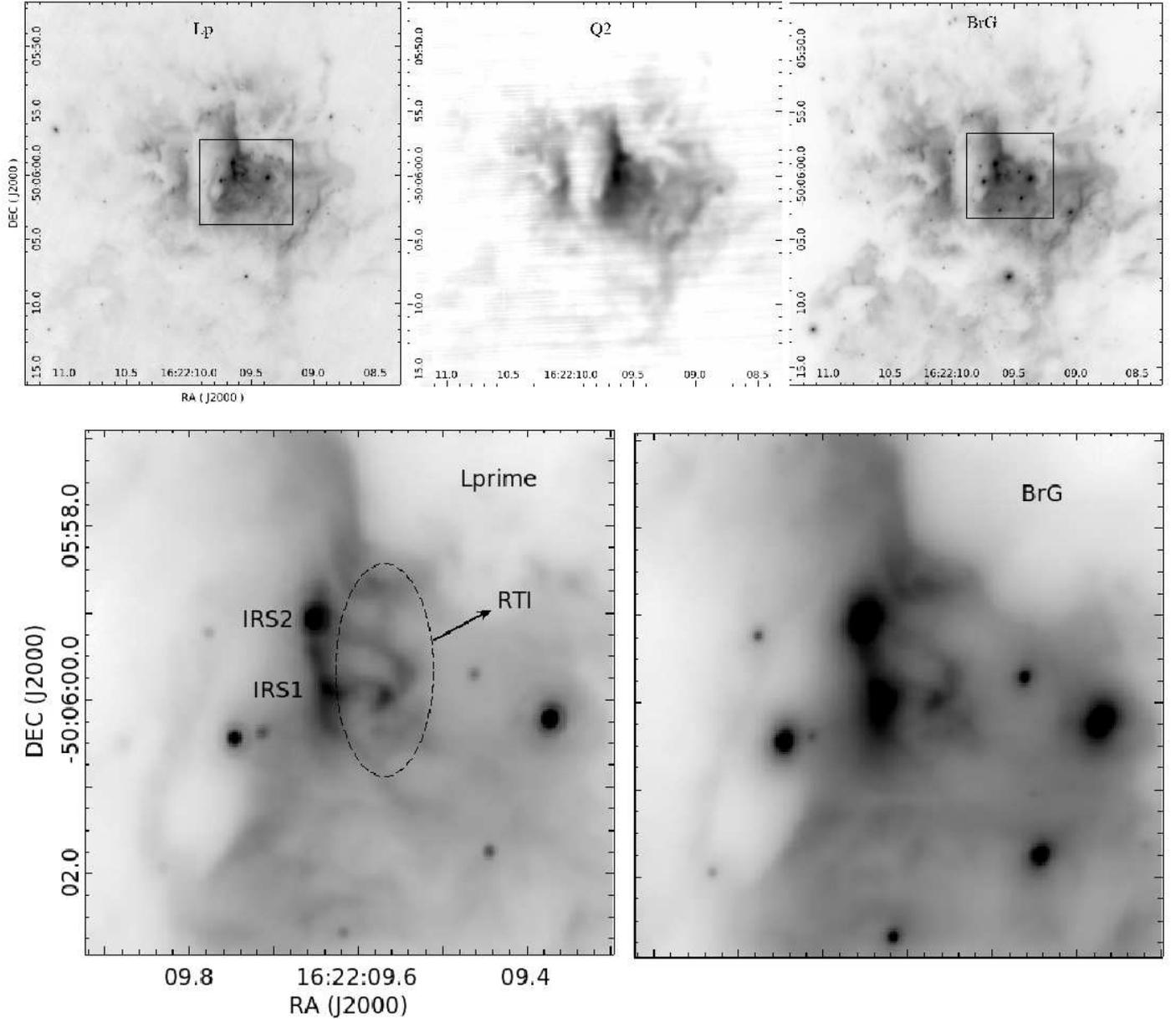}}
      \caption{ Top panel: The NACO and VISIR data sample in the
        L$^{\prime}$, Q2, and Br$\gamma$ bands of the G333.6-0.2 UCHII
        region is displayed using a logarithmic inverted grey scale. The
        images display noticeable differences from the near-infrared
        to mid-infrared, which can be better visualized from
        Fig.\,4a. Bottom panel: A zoom-in view of the central region
        (marked with a box in the top panel) displayed using square
        root scaling to show the IRS sources and RTI features
        clearly. }
         \label{images}
   \end{figure*}

\begin{figure*}
   \centering

 \resizebox{\hsize}{!}{\includegraphics{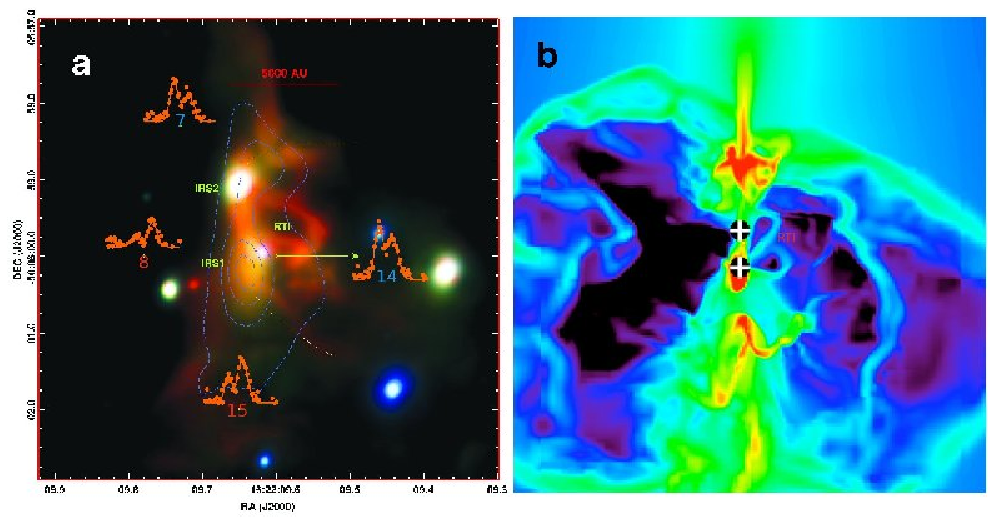}}
  \caption{a) Near-infrared H-, Br$\gamma$-, and L$^{\prime}$-band
    adaptive optics data coded in blue, green, and red respectively
    to produce the colour composite of the brightest part of the
    ultra-compact HII region G333.6-0.2. Blue dashed contours depict
    18$\mu$m emission revealing the source IRS\,1.  H90$\alpha$
    recombination-line double-peaked spectra from \cite{fujiyoshi06}
    are overplotted. The numbers below these spectra correspond to the
    numbering in the original paper (see their Fig.4 and 5). The thin
    orange line sketch represents possible margin of the outflow
    bubble. b) Simulations of a forming massive star system from
    \citet{mk09} showing 3000~AU$\times$3000~AU region at 41700~years,
    where the colours represent volume densities. The disk plane and
    outflow axis are aligned vertically and horizontally.}
         \label{images}
   \end{figure*}


\section{Characteristics of the G333.6-0.2 region}
A near-infrared K$_s$-band view of the G333.6-0.2 region can be found
in Fig.~1. At a seeing of 0.5\arcsec, this image is an improvement
over similar images obtained with a 4~m class telescope described by
\citet{fujiyoshi05}. Mapped at the resolution of an arcsecond
\citep{fujiyoshi06}, the UCHII region is fairly accurately represented
by the bright near-infrared bipolar-shaped nebula, separated by a dark
patch seen in Fig.~1. The dark patches delineating the midriff of
bipolar-shaped UCHII regions may represent large molecular toroids, a
classical example being S106 region in the Cygnus
region. High-resolution 1.3cm mapping of S106 region reveals a
well-defined bipolar-shaped HII region \citep{felli84}. The optical
and near-infrared dark patch at the midriff of the S106 compact HII
region is found to represent a large toroid traced by 2mm dust
continuum emission \citep{furuya02}. Similarly, the dark patch here in
G333.6-0.2 may represent a large toroid; uncovering it will require
high-spatial-resolution observations with appropriate molecular-line and
dust tracers.

The margin of the western lobe of the bipolar nebula in G333.6-0.2 is
so bright that it appears saturated in most infrared images obtained
even with moderate facilities. This is because two ionizing sources
are located at this margin (and at the edge of the dark patch): one
bright at mid-infrared and the other at near-infrared
wavelengths. They are denoted as IRS\,1 and IRS\,2 (see Figures 3 and
4a). The brighter of the two peaks at longer wavelengths, IRS\,1
becomes increasingly bright from 12~$\mu$m to 18~$\mu$m, resolving out
as a clear elliptical structure in the 18~$\mu$m\, image (blue dashed
contours in Fig.\,4a). This elliptical structure represent polycyclic
aromatic hydrocarbon and warm dust emission arising from a
circumstellar disk/envelope. The NACO-AO data shows that IRS\,1 is
invisible at shorter wavelengths, whereas IRS\,2 appears as a star.
Indeed, the 18$\mu$m emission surrounds IRS\,2 rather than
peaking on it. The low-intensity peak coincides with a small
dark lane to the northwest of IRS\,2, as can be seen from the overlays
in Fig.\,4a. The sources IRS\,1 and IRS\,2 therefore appear to be
embedded in a single, large toroidal structure of $\sim$5000\,AU in
diameter seen edge-on. The J,~H, and K$_s$ magnitudes of IRS\,2 are 13.1,
11.7 and 10.2 respectively.  The faint near-infrared star close to
IRS\,1 (Fig.\,4a) is a reddened object without any infrared excess,
the nature of which is difficult to classify in the absence of an
accurate extinction measurement of the region probed. The 10\%
accuracy of adaptive optics photometry is insufficient to understand
the accurate nature of these sources, particularly in this heavily
extincted region. The UCHII region itself is modelled to represent a
champagne-flow-like outflow cavity bubble expanding in a perpendicular
direction to the plane of this toroidal structure hosting the two
sources \citep{fujiyoshi06}. The ionised outflow cavity is traced and
characterised by the H90${\alpha}$ radio recombination and centimeter
continuum imaging at 1.6\arcsec\, resolution \citep{fujiyoshi06}.

\subsection{Filaments and finger/hook like features}

From Figures.3 and 4a it can be seen that tiny filamentary features
with finger/hook morphology are found along the proposed outflow
cavity. We note that they appear prominently amidst bright nebular
emission, which is likely swamped by reflected light. At a spatial
resolution approximately ten times better, these features represent
details of previously known morphology \citep{fujiyoshi01}. The
8-12$\mu$m spectro-polarimetric studies have been used to show that
the emission in this region is composed of both absorptive and
emissive polarized components \citep{fujiyoshi01}. Also, significant
line emission in the Br$\gamma$ and 12.8$\mu$m [NeII] emission
originates from this compact region
\citep{fujiyoshi98,fujiyoshi06,grave13}.  Therefore the filamentary
features are very likely tracing layers of hot ionized gas and/or dust
with predominant small particle emission. Given the high temperatures
in the close vicinity of the luminous sources, larger grains are
typically destroyed. Similarly, molecular gas should be completely
absent by exclusion effect of the above primary composition; this can
be justified to a certain degree by the lack of H$_2$ emission in this
region \citep{mm89,fujiyoshi06}. It is to be noted that the existing
simple narrow-band imaging failed to produce discernable features
because continuum subtraction is very uncertain and not effective in a
region such as this without spectral imaging. Nevertheless,
indications for such properties can be visualised through a colour
composite, as shown in Fig.4a. Because the L$^{\prime}$-, Br$\gamma$-,
and H-band images used in the colour composite are all scaled
similarly, the red and green colours represent relatively denser and
more ionised parts. These spatially resolved features are limited to
$<$5000AU in length. Such small-scale filamentary loops have not been
found in surveys of UCHII regions or compact nebulae.  The length and
width of the most prominent filaments are measured on the L$^{\prime}$
image at various positions. The lengths are measured using a straight
line and the widths by using intensity profile cuts at multiple
positions across the filaments. These measurements result in mean
values of 0.63\arcsec$\pm$0.29\arcsec and 0.16\arcsec$\pm$0.03\arcsec
for the lengths and widths of the filaments respectively. At a
distance of 3.6\,kpc, this estimate translates to a mean length of
2268$\pm$1044\,AU and mean width of 576$\pm$108\,AU. The observed
widths ($\sim$500AU) are mostly an effect of the 0.1\arcsec\, spatial
resolution from the adaptive optics images. The longest filament is
the one associated with IRS\,2 source measuring $\sim$4000\,AU in
length.  Filamentary structures in the interstellar medium (ISM) are
known to be ubiquitous in the Galaxy and are found in all molecular
clouds. The ISM filaments are much larger in size: spatially resolved
characterization using HERSCHEL data estimate their widths to be
0.1$\pm$0.03\,pc $\sim$ 20000\,AU \citep{arz11}, which is a factor of
four larger than the {\it lengths} of the features found here.

\section{Radiation Rayleigh-Taylor Instabilities?}

The filamentary features found above were exhaustively compared with
the outcome of numerical simulations from KKM09, \citet{ajc11},
\citet{hen11}, and \citet{peters10a,peters10b,peters11}, which have
produced structures at a few hundred to a few thousand AU. Apart from
comparing the morphology, the observed properties in the infrared to
radio are analysed and supplemented by simple calculations. Towards
the end of this section, it becomes apparent that the net result
of this analysis suggests that the observed filamentary features likely
represent RTI features residing in the ionized outflow cavity of
G333.6-0.2. Indeed, as shown in Figs.\,4a and 4b, the core of
G333.6-0.2 displays good resemblance with a snapshot of the edge-on
view of the simulations by KKM09. In the following, I present
evidence to show that the observed resemblance between data and
simulation snapshot is more than simple coincidence.

To begin, I recall the salient characteristics of the sub-structures
produced by the simulations under consideration: a) filaments due to
RTI prominently appear in the outflow cavity in the simulations of
KKM09 and are essentially absent elsewhere, b) filaments produced by
\citet{hen11} are radially oriented, and appear in all three axes
(x,y,z) representing magnetised ropes of material converging to a
centre, and c) simulations of \citet{peters10a,peters10b,peters11}
result in small structures caused by the pressure of photoionized gas,
which typically arises outside a region where the escape speed is
$>$10\kms\,.

Figure.\,4a display two salient features: a) filamentary features with
finger/hook morphology that are {\em sharply limited} in size to
$\le$1.3\arcsec and also appear connected to IRS\,1 and IRS\,2, b)
sources IRS\,1 and IRS\,2 appear similar in mass, despite their
different infrared properties because the temperature estimate is
roughly the same for both sources at $>$40000K \citep[see][]{grave13}.

As evidenced from NACO data in Fig.\,3 and 4a, the features in question
are restricted to a small space around the forming stars and limited
in lengths, preferentially found within the previously known outflow
cavity. Therefore they are unlike the radially converging magnetised
rope-like filaments from \citet{hen11}. Next, the short lengths and
visibly linked nature to the IRS sources show that they arise from
within a region where the escape speeds are greater than 10\kms (see
below).  The smallest and brightest finger/hook--like feature is
$\sim$2000AU in length and appears connected to the IRS
sources. The escape velocity for photoionised gas at this distance is
$\sim$8 km s$^{-1}$, considering a star of 40\msun\, surrounded by a
disk of 20\msun. In reality, the gravitational pull will be much
larger because: a) together, IRS\,1 and IRS\,2 can contribute a mass
$\sim$100\msun\, or more, and b) the disk/envelope plane will contain
an even larger parcel of mass. This assumption is made considering the
elliptically shaped hot dust emission at 18$\mu$m, which is primarily
due to small-grain dust (micron sized) and also the spatial
coincidence with the 1.2\,mm peak representing the massive and dense
clump (15~000\msun\, and n$_{H_2}$ = 5.44$\times$10$^{4}$~cm$^{-3}$)
\citep{bm04}. Also, the observed features are not due to hot gas
expanding at the sound speed of $\sim$10~\kms\,. This is because the
sound speed and the escape velocity are comparable in magnitude, making
it improbable that hot gas pressure drives the 2000-3000 AU large
observed features. Indeed, as elaborated in Sec.\,5, the features
coincide with a region where double-peaked H90$\alpha$ recombination
lines clearly display peak-to-peak velocities of 20-30\kms\,. 
  These double peaks are resolved components located on opposite sides
  with respect to the ambient velocity, which is found to be V$_{LSR}$=
  -47\kms\, by lower spatial resolution observations \citep{McGee75}.
Together, these considerations suggest that the observed features are
unlikely to be those arising in the simulations by
\citet{peters10a,peters10b,peters11} and are also not due to hot gas
expanding at the sound speed of $\sim$10~\kms\,.

Instead, the observed features satisfy many qualities expected from
RTI structures produced by KKM09, as discussed below. The first
requirement for the development of RTI is that radiation (isotropic)
should be strong enough to imitate a light fluid {\em capable} of
upholding a denser overlying fluid. The two sources IRS\,1 and IRS\,2
are the only resolved ones that coincide with the peak of the UCHII
region G333.6-0.2, which is known to emit LyC photons equivalent to
that of 19 O7V stars. These sources are $\sim$1\arcsec\, apart,
corresponding to $\sim$3600\,AU at a distance of
3.6\,kpc. Mid-infrared spectra obtained with the VISIR instrument on
VLT at a similar spatial resolution of 0.3\arcsec\, by placing the
slit along IRS\,1 and IRS\,2 have uncovered intense emission of
[ArIII] and [NeII] lines, which localise and peak (within 0.3\arcsec)
on these sources. Radiative transfer analysis of the line fluxes
resulted in calculating high temperatures of $\sim$40000-50000~K. It
led \citet{grave13} to argue that the spectral types of the embedded
stars were $\sim$O3-O5 \citep{grave13}.  The temperature estimate is
uncertain by up to 5000K, depending on other factors . Even so, the
two sources IRS\,1 and IRS\,2 can account for at least half of the
total bolometric luminosity and most of the LyC flux estimated for the
G333.6-0.2 UCHII region. We note that \citet{rubin94} estimated a
temperature of $\sim$36000\,K for this region, even when they used
far-infrared spectroscopic data obtained with beam sizes nearly a
factor of ten larger than those used by \citet{grave13}. These data
therefore included contributions from other competing ionizing sources
in the vicinity.  The resulting radiation flux from IRS\,1 and IRS\,2
is large enough to imitate a light fluid, if appropriately reprocessed
and redistributed.

I will present a simplistic analysis of various pressure contributions
similar to the analysis carried out by \citet{lopez11} to compare it
with the simulations of KKM09. Such analysis is useful to study the
global energy budget in HII regions, and computing forces on an
irradiated medium requires a detailed treatment similar to that of
\citet{jk11} and \citet{jiang13}. The choice to nevertheless adopt the
method of \citet{lopez11} is motivated by the limited data on hand,
which lack sufficient measurements on opacity (and its variation), as
well as the mass and composition of the material in the region. This
analysis is meant only to demonstrate that the radiation pressure
supercedes the ionized gas pressure in a general sense. If we assume
that both IRS\,1 and IRS\,2 represent on average O5.5V spectral-type
objects (a conservative adoption based on the temperature mentioned
above), following \citet{vacca96}, each of these sources should have a
luminosity of L$_*$$\sim$4$\times$10$^5$~\lsun\,. The direct radiation
pressure can then be calculated using the equation P$_{dir}$ = L /
4$\pi$r$^2$c, under the assumption that optical depth $\tau
\sim$1. This pressure is 1.4$\times$10$^{-7}$~Pa and
3.86$\times$10$^{-9}$~Pa respectively at distances of 1\arcsec and
6\arcsec (3600~AU and 21000~AU) from the plane joining the two
sources. It is comparable to the ionized gas pressure of
11.5$\times$10$^{-8}$~Pa, which is computed by adopting the electron
number density and temperature estimated from H90$\alpha$ observations
\citep{fujiyoshi06} and using the equation P$_{HII} \sim $2n$_e$
kT$_{HII}$. Because the spatial resolution of the H90$\alpha$
observations is 1.6\arcsec, we have used the values n$_e$=513000
cm$^{-3}$ and T$_{HII}$=8350~K, which were estimated using the data
closest to the plane of IRS\,1 and IRS\,2.

The pressure from the X-ray emitting gas, which includes contributions
from the stellar winds, was calculated using the equation {\em
  P$_X$ = 1.9n$_X$kT$_X$}, where n$_X$ and kT$_X$ were obtained from
the Chandra X-ray data analysis (see Sec.~2). The resulting pressure
is about two to three orders of magnitude lower than the radiation
pressure and therefore considered insignificant.

These conservative ``back of the envelope'' calculations show that in
this particular target, the radiation pressure supercedes the ionized
gas pressure at least up to a distance of 5000AU from the plane
containing IRS sources. The direct radiation can behave like a light
fluid only when the radial isotropy is destroyed; this can happen primarily
through turbulence within the outflow bubble and/or the
irregular dust envelope evidenced by the 18$\mu$m data, which can
reprocess and re-radiate through scattering events. The resulting
non-isotropic flux escaping through the optically thin ionised outflow
bubble is then powerful enough to imitate a lighter fluid upholding
the dense overlying fluid, which in this case can be largely ionized.

I note the concentration of features in Figs.~3 and 4a at about
1\arcsec\, above the plane of the sources. This is approximately the
distance up to which the pressure from the radiation fluid balances
the heavier fluid above. When this delicate balance is perturbed,
which can occur primarily due to the turbulence within the ionized
gas, RTI can develop.  In this case, it is visualised by the observed
features. The instabilities result in dense structures, which are then
pulled by the gravitational force towards the disk plane. The two
linear features with finger/hook morphology in Figs.~3 and ~4a, which
appear connected to the IRS sources, then represent such pulled-back
structures.

In this simplistic scenario, the radiation pressure supercedes
gravitational pressure beyond a certain distance from the star and
equals that of ionized gas pressure (about 5000\,AU in this case),
which can explain the balancing zone by the lighter radiation fluid of
the overlying dense material fluid. It should be noted that in the
treatment of generalized Eddington limit, such effects would not be
possible because both the radiation and gravitational forces scale as
r$^{-2}$. It is expected to occur only when a high light-to-mass ratio
(L/M $\sim$ 2500), driven by Kelvin-Helmholtz contraction, is
sustained during the early stages. The requirement here is that the
radiation be dissipated in a preferential direction, as discussed in
KKM09. In this particular target, IRS\,1 and IRS\,2 have multiple
characteristics of representing single massive stars, as discussed by
\citet{grave13}. The hypothesis is based on the statistically rare
case of tight, very massive equal-mass binaries with separations of
less than 500AU \citep[for e.g.][]{sana12}. When the mass, luminosity,
and multiplicity of the two IRS sources are ascertained, it will be
possible to accurately compute the pressure balance using future
observations.

Additionally, in the scenario of KKM09, the infalling gas strikes the
walls of the radiation-driven bubble, which will slide back along the
walls to the mid-plane. The two outermost vertical structures seen in
Fig.~4a (marked using orange lines) can represent such walls of the
radiation bubble, curiously, no dense features are observed beyond
these walls.

\section{Discussion}

The density structures discussed above, which are satisfying
signatures of RTI, should be overdensities of ionized gas as it is
completely embedded inside the observed compact HII region.  The
conjecture made in the previous section that the observed features are
not due to hot gas expanding at the sound speed of $\sim$10~\kms\, is
supported by the observed H90$\alpha$ data \citep{fujiyoshi06}. Of the
36 H90${\alpha}$ profiles obtained to map this target, six are clearly
double peaked; the two peaks are found on either side of V$_{LSR}$,
and have an average peak-to-peak velocity shift of 30~\kms\,. These
double-peaked profiles closely overlap and surround the region
containing the newly discovered density features. In Fig.~4a, four of
the most prominent profiles are overplotted on the adaptive optics
image, which can be seen to surround the sources IRS\,1 and
IRS\,2. This overlay allows an ad hoc interpretation of the previous
H90${\alpha}$ data, proving which demands higher resolution radio
data: a) profiles along the east-west direction originate in an
expanding outflow bubble (the two peaks are due to the two walls of
the bubble) and b) profiles along the north-south direction represent
some rotation involving two components.

The bipolar nebula separated by a dark lane ( as depicted by Fig.~1)
is an indication that the system is viewed at a small inclination
angle in the sky-plane, with the disk/envelope appearing nearly
edge-on.  As the outflow axis is nearly aligned in the plane of the
sky, the two peaks (blue and red shifted) of the double-peaked
profiles represent the average motion of matter moving toward and away
from us within the expanding bubble. The line-of-sight velocity
(peak-to-peak difference) depicted by H90${\alpha}$ profiles \#8 and
\#14 \citep[see][]{fujiyoshi06} of 20~\kms\, and 24~\kms\, correspond
to these components in the two outflow lobes.

The profiles \#7 and \#15 which roughly enclose the two edges of the
observed infrared dark lane, can represent rotation of a large
($\sim$5000~AU) toroid.  The double peaks and the shift in the heights
of blue and red peaks arising from opposite sides can be explained
only if this rotating structure includes two individual structures
rotating in the same direction and viewed edge-on. Such a scenario
strikingly resembles the simulations by KKM09. Given that the
H90$\alpha$ observations are obtained at a spatial resolution of
1.6\arcsec\,, motions at smaller scales are mixed up and averaged out,
resulting in a double peak at the two edges.  I note that
\cite{fujiyoshi06} considered the hypothesis of rotation and rejected
it to propose blister/champagne flow to explain the double-peaked
profiles. This interpretation was understandably influenced by the
uncertainty that the heart of G333.6-0.2 hosts a cluster of O7-O8
ionizing sources. The new observations here and in \citet{grave13}
pinpoint IRS\,1 and IRS\,2 as the main luminous sources resulting in
the high L/M condition necessary for the proposed scenario. Indeed,
the proposed outflow interpretation of \citet{fujiyoshi06} can now be
associated directly with IRS\,1.  In hindsight, it also appears that
the dark lane visible in the near-infrared images may not be a toroid
after all. The real toroid or disks must coincide with the IRS sources
and the 18$\mu$m emission contours, while the dark patch marked as a
possible toroid in Fig.\,1 is actually a result of low-density cavity
material. This can be visualised by direct comparison of Fig.\,4a and
4b.

Essentially all features shown in Fig.~4a, namely the disk/toroid and
the outflow cavity encompassing the observed sources, are immersed and
depict details of the compact HII region.  If the above ad hoc
interpretation of the H90$\alpha$ double-peaked profile is proven to
be valid, the sources IRS\,1 and IRS\,2 are likely to be accreting ionized
gas (rather than molecular material), as proposed by
\cite{keto02,keto03}.  In fact, H$_2$ emission is found to be
insignificant in this region \citep{mm89,fujiyoshi06}, arguing for the
lack of any molecular gas.

If the observed compact HII region \citep{fujiyoshi06} in this target
($\sim$6\arcsec$\sim$20000~AU) represents the radiation-driven bubble,
then it appears to be much larger than the $\sim$3000AU bubble from
simulations by KKM09. It is possible that the observed 6\arcsec size
of the compact HII region has significant contributions from other
competing sources in the vicinity. Also, the sizescale of RTI
predicted by KKM09 is a factor of five to ten smaller than those
discovered here. The uncertainties inherent to both observations and
theory impede a quantitative comparison at this stage. Next, the KKM09
simulations did not consider the influence of factors such as direct
radiation, magnetic fields, and photoionisation, which can impact the
size scales of the expected features. A detailed discussion on the
effects of various caveats in producing RTI is presented by
\citet{kt12}. While many arguments are used here to discuss the less
likely role of such other factors, ruling out their contribution
against RTI can be challenging, especially because simulations
including them also resulted in producing fine filamentary features
\citep{peters11}. A multitude of new high-spatial resolution
measurements evaluating the composition, opacity, mass, luminosity
etc., should be quantitatively compared to the simulation outcomes
\citep{jk11,jiang13,kt12} before confirming the crucial role of RTI in
the mechanism of high-mass star formation. The observations and
analysis presented here nevertheless show that RTI may indeed
play an important role in the formation of massive stars and
observational signatures at physical scale resolutions of a thousand
AU or less will require careful attention in the future to confront
theoretical scenarios.

\begin{acknowledgements}

I am grateful to Mark Krumholz for inspiring and educative discussions
of remarkable clarity, which enlightened me with the essentials of
some theoretical investigations, that had been beyond my grasp by solely reading
the literature. This work greatly benefitted from the objective
criticism and input from anonymous referees, both for the present
manuscript and from earlier submissions to Nature and Science
journals. Warm thanks are due to A.R.Rao of TIFR, Mumbai for his
advice on the X-ray data analysis and to outreach team at CAUP for help
with Figure.~3. Part of this work was carried out under the auspices
of an EU FP7 Marie-Curie IRSES-230843 programme. I am supported by a
Ci\^encia 2007 contract, funded by FCT (Portugal) and POPH/FSE
(EC). This work has made use of ESO archival services, Spitzer, and
Chandra Data Archives.

\end{acknowledgements}

\bibliographystyle{aa}

\end{document}